\documentclass[vecphys]{svmult}

\usepackage{makeidx}         
\usepackage{graphicx}        
\usepackage{multicol}        
\usepackage[bottom]{footmisc}

\makeindex             

\begin{document}

\title*{Cosmological parameters from Galaxy Clusters: an Introduction}
\author{Paolo Tozzi}
\institute{INAF -- Osservatorio Astronomico di Trieste, via
G.B. Tiepolo 11, 34131 Trieste -- Italy.  email:
\texttt{tozzi@ts.astro.it}}
%
%
\maketitle

This lecture is an introduction to cosmological tests with clusters of
galaxies.  Here I do not intend to provide a complete review of the
subject, but rather to describe the basic procedures to set up the
{\sl fitting machinery} to constrain cosmological parameters from
clusters, and to show how to handle data with a critical insight.  I
will focus mainly on the properties of X--ray clusters of galaxies,
showing their success as cosmological tools, to end up discussing the
complex thermodynamics of the diffuse intracluster medium and its
impact on the cosmological tests.

\section*{Introduction}
\label{intro}

This lecture concerns a classic topic of observational and theoretical
astrophysics: investigating the global properties of the Universe by
looking at its large scale structure.  In particular, we are
interested in using our knowledge on the physical properties of
clusters of galaxies and their distribution with mass and cosmic epoch
to put constraints on the cosmological parameters, namely the matter
density parameter $\Omega_0$ and the dark energy component (parameter
$w$ or, in the simplest case $w=-1$, the cosmological constant
$\Lambda$).  Our journey will be a round trip: starting from a simple
theoretical approach, we will build a powerful tool to interpret the
data and measure the cosmological parameters, but then, we will be
forced to go back to theory for a more complex approach to the physics
of clusters of galaxies.

The theoretical starting point (Section \S 1) provides a reasonable
framework to understand the formation and evolution of clusters, which
are the most massive bound and quasi--relaxed objects in the Universe,
in a cosmological context.  The observational part (Section \S 2) will
focus mostly on X--ray observations, which offered the most important
observational window for this kind of test for the last 15 years.  As
often happens in astrophysics, we will find that the increasing
quality of the data sheds light on a situation much more complex than
previously thought.  The most recent data, collected in the last five
years by the Chandra and XMM--Newton satellites, calls for a much
deeper understanding of the physics of baryons in clusters of
galaxies, forcing us to reconsider the basic physical ingredients to
make a more robust connection between clusters and cosmology (Section
\S 3).  This effort is worth, since clusters are an invaluable tool
for cosmology, and they can significantly constrain the cosmological
parameters in a way which is complementary to the other classic
cosmological tests (the Cosmic Microwave Background, hereafter CMB,
and Type Ia Supernovae, SneIa).

\section{Clusters of galaxies in a cosmological context}

\subsection{What is a cluster of galaxies}

We start with a simple definition of what is a cluster of galaxies.
The simplest approach is to identify a cluster as an overdensity in
the projected distribution of galaxies in an optical image. The first
catalog was indeed a compilation of galaxy concentrations found by eye
in optical images (Abell 1958).  Today, the quality of optical images,
especially that from the Hubble Space Telescope, are such that bright
(or, in terms of galaxies, rich) clusters of galaxies are among the
most spectacular objects of the extragalactic sky.  In Figure
\ref{a1689}, first panel, we show an optical image of Abell 1689, a
massive cluster at redshift $z=0.18$.  The bright, yellowish galaxies
are the massive ellipticals which typically populate the inner part of
rich clusters.  In this image it is also possible to see background
galaxies distorted by gravitational lensing.

However, the stars in the cluster galaxies, visible in the optical
light, are not at all the dominant mass component. The X--ray image of
Abell 1689 obtained with the Chandra satellite (second panel of Figure
\ref{a1689}) shows the distribution of hot gas, which is the dominant
baryonic component. The total mass, anyway, is dominated by the
non--baryonic component called dark matter (see the reconstructed
distribution in the third panel).  To review the properties and the
hypothesis on the nature of the dark matter see Kolb \& Turner (1990).
Here we need to know only that dark matter is collisionless and that
it dominates gravitationally large objects like groups and clusters of
galaxies.

\begin{figure}
\centering
%
\vskip 1.5cm
\begin{center}
{\bf see fig1.gif}
\end{center}
\vskip 1.5cm
\caption{The rich cluster of galaxies Abell 1689 ($z=0.18$).  The
three images show the three main components in terms of mass.  In
order of increasing mass fraction, from left to right: an optical
image (stars) taken with the Hubble Space Telescope (credits ACS
Science Team, ESA NASA); an X--ray image taken with Chandra (showing
the diffuse Intra Cluster Medium); the dark matter map reconstructed
from lensing (after Broadhurst et al. 2005).}
\label{a1689}     
\end{figure}

To be more quantitative, the composition of a cluster of galaxies is
roughly as follows: 80\% of the mass is in dark matter; 17\% in hot
diffuse baryons, the so--called IntraCluster Medium (ICM); 3\% in the
form of cooled barions, meaning stars or cold gas.  The total mass of
clusters ranges from few $\times 10^{13} M_\odot$ (small groups) to
more than $10^{15} M_\odot$.  While the baryonic components can be
directly observed (mainly in the optical, infrared and near infrared
bands for the stars and in the X--ray band for the ICM), the dark
matter can be measured only through the effect of gravitational
lensing on the background galaxies or by other dynamical properties of
clusters.  Needless to say, the total mass of a cluster is the
fundamental quantity we need to know.  A useful definition of the
dynamical mass of a cluster will be given after briefly discussing the
physics of gravitational collapse.

\subsection{The linear theory of gravitational collapse}

Clusters form through gravitational collapse, which is driven by dark
matter.  This is strongly simplifying our problem, since the dark
matter, whatever it is, must behave as a collisionless fluid, and
therefore it is not affected by dissipative processes, unlike the
baryons, which are pressure supported, and experience radiative
cooling.  Since we are interested in the total mass, we can neglect,
on a first instance, the physical processes affecting only the
baryons.

To describe the evolution of a collisionless fluid under its own
gravity, we can use the eulerian equations of motion describing a
perfect fluid assuming spherical symmetry (continuity, Euler and
Poisson equations, see Kolb \& Turner 1990):

\begin{eqnarray}
{{\partial \rho}\over {\partial t}} + \vec{\nabla} \bullet (\rho \, \vec
v ) = 0 \\
{{\partial \vec v}\over {\partial t}} + (\vec v \bullet \vec \nabla )
\vec v + {1\over \rho} \vec \nabla p + \vec \nabla \phi = 0 \\
\nabla ^2 \phi = 4 \pi G \rho \, ,
\end{eqnarray}

\noindent
where $\rho$ is the density field, $\vec v$ is the velocity field, $p$
is the pressure and $\phi$ is the gravitational potential generated by
the density field itself.  We are interested in how the density
evolves with time.  First, we consider small positive density
perturbations with respect to a uniform and static background with
density $\rho_0$, so that we can easily linearize the system of
equations.  We define our interesting variable as the overdensity
$\delta \equiv (\rho - \rho_0)/\rho_0$, and assume that the
unperturbed solution is a static background, $\rho = \rho_0 =
const$\footnote{This is not correct since the Poisson equation is not
satisfied; however this assumption, called the {\sl Jeans swindle},
leads to correct consequences. }.  We just need a little algebra to
linearize the equations and derive the solution for the density
contrast in terms if its linear components $\delta_k = A \, \tt exp
[-i\vec k \vec r + i \omega t]$.  After defining the sound speed as
$v^2_s \equiv ({{\partial p}\over {\partial \rho}})_{adiabatic}$, the
solution can be written as follows:

\begin{equation}
\ddot \delta_k  = (4\pi G\rho_0 - v^2_s k^2) \delta_k \, ,
\end{equation}

\noindent
or, in a very familiar way, $\ddot \delta = -\omega^2 \delta$.  The
solution is therefore an armonic oscillator with dispersion relation $
\omega^2 = v^2_s k^2 - 4\pi G\rho_0$.  Note that for dark matter,
$v_s$ is substituted by the velocity dispersion of the collisionsless
particles $v_*$.  When $\omega^2$ is negative, the solution behaves
exponentially.  This qualitative result is largely expected in this
extremely simplified situation: in a static background, the
gravitational force is proportional to the overdensity itself, and the
gravitational instability evolves rapidly.  The dispersion relation
defines a length scale $\sim 1/k$ for which the perturbation is
unstable.

However, we are interested in the realistic solution in an expanding
background.  This can be obtained by substituting a varying background
density $\rho_0 = \rho_0(t_0) R^{-3}(t)$ in the equations, where
$R(t)$ is the scale factor satisfying the usual Friedmann equation.
The expansion of the Universe is conveniently expressed trough the
fractional growth of $R(t)$ which is the Hubble function $H(t) = \dot
R/R$.  The solution of the linearized problem satisfies:

\begin{equation}
\ddot \delta_k + 2 {{\dot R}\over R} \dot \delta_k + \Big( v^2_s k^2 -
4\pi G \rho_0 \Big) \delta_k = 0 \, .
\label{deltaevol}
\end{equation}

\noindent
The additional term $2 {{\dot R}\over R} \dot \delta_k$ changes
considerably the qualitative behaviour of the solution, depending on
the behaviour of $R(t)$.  To show a specific example, we adopt
$R\propto t^{2/3}$, or $\dot R/R = (2/3)(t/t_0)^{-1}$, appropriate for
an Einstein--de--Sitter Universe (EdS, $\Omega_0 = 1$), to find:

\begin{equation}
\ddot \delta_k + {4\over {3t }} \dot \delta_k - {2 \over {3 t^2}}
\delta_k = 0 \, 
\end{equation}

\noindent
(note that here we assumed a negligible $v_s$ or $v_*$ as appropriate
for Cold Dark Matter).  The growing mode solution is $\delta_+(t) =
\delta_+(t_i) (t / t_i) ^{2/3}$.  Therefore, in an EdS Universe, we
have the remarkably simple result that the linear growth of a density
perturbation is proportional to the expansion factor $ (1+z)$.  One
may wonder why we show the solution for $\Omega_0=1$, while we are
here in this School to learn that dark energy is the dominant
component in the Universe, while the matter component is $\Omega_0
\leq 0.3$.  The fact is that the case for $\Omega_0=1$ gives simple
analytical solutions, an occurrence that contributed substantially to
the succes of the EdS Universe untill the early 90s, when
observational evidences started to point towards a low matter density,
making room for the debout of dark energy.

More in general, we find that the fastest is the expansion, the
slowest is the linear growth of perturbations.  The link between the
expansion rate of the Universe and the rate of collapse of density
perturbations is strongest at the largest scales.  This is because
large--scale perturbations are the last to leave the linear phase,
while smaller scales (the one from which galaxies form, for example),
collapsed earlier.  This, in turn, is a consequence of the shape of
the primordial spectrum of the density perturbations, and it is true
in any cold dark matter (CDM) dominated Universe.  We will discuss
this aspect in greater detail later.

\subsection{Non linear evolution of density perturbations and
virialization}

Now we have a simple framework which allows us to compute the linear
phase of collapse of a spherical density perturbations in an expanding
universe.  However, our final goal is to describe clusters of
galaxies, which are definitely non linear (and non--spherical, but
spherical symmetry is too convenient to be dropped!).  In addition, we
need to define accurately the total dynamical mass of a relaxed
object.  Should we abandon the simple linear treatment to look for a
more complex and computationally heavier approach?  Luckily for us, we
can define a relaxed object in terms of the same parameters entering
the linear theory, as shown in the following pages.

Thanks to the Birkhoff theorem, we can ignore what is outside a
perturbation, and we can describe a uniform spherical ({\sl top--hat})
perturbation like a sub--universe with a density larger than the
critical one $\Omega \geq 1$.  Such a universe would expand and
recollapse in a finite time.  If we consider a spherical shell
encompassing the overdensity, we can use the Friedmann--Robertson
Walker (FRW) model for the evolution of each shell in a
parametric form:

\begin{equation}
R = {{GM}\over {2E_0}} \Big(1-{\tt cos}(\eta)\Big) \,,  \, \, \, \, \, \,
t = {{GM}\over{(2E_0)^{3/2}}} \Big(\eta -{\tt sin}(\eta)\Big) \, .
\end{equation}

The maximum of the expansion radius defines the turn--around time,
which is the epoch when the shell starts to collapse, after decoupling
from the cosmic expansion.  Due to the simmetry of the solution, the
time of collapse is twice the turn--around time.  In our spherical
approximation, the collapse ends into a singularity.  What is actually
happening to a real, non--spherical perturbation, is that the
different shells cross each other and start oscillating across the
center.  However, we can bravely assume that, by that time, the
perturbation (meaning all the mass included in the outermost spherical
shell) is evolved into a spherical, self--gravitating {\sl virialized}
halo.

A virialized halo is a region of space where matter is gravitationally
bound, and where a statistical equilibrium between the potential and
the kinetic energy is established.  Every mass component participates
to the equilibrium: both the diffuse, ionized gas, the galaxies, and
the dark matter particles, have random velocities described by a
maxwellian distribution with the same temperature.  The virialization
condition in its simplest form reads $2T+U=0$, where $T = M_{tot}
\langle v^2\rangle /2$ is the average kinetic energy per particle, and
$U = -G M^2_{tot}/R_c$ is the average potential energy.  Energy
conservation argument fix the relation between mass and the
characteristic radius $R_c$ of the halo, so that the virial theorem
effectively establish a one--to--one correspondence between the total
dynamical mass and the virial temperature.

\begin{figure}
\centering
\includegraphics[height=6cm]{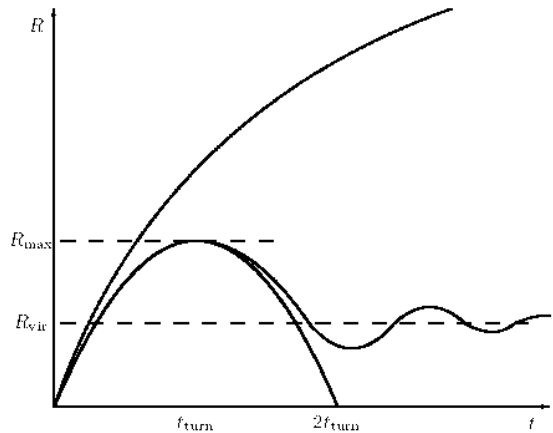}
\caption{The evolution with time of a {\sl top--hat} perturbation.
The upper curve is the expansion of the exterior mass shell, while the
closed curve is the solution which behaves like a closed FRW model.
The wavy curve is the radius of a realistic perturbation which bounces
back and virializes after few oscillations (from Kaiser 2002).}
\label{sphcoll}     
\end{figure}

Going back to the linear solution, how can we describe the formation
of virialized halos in terms of the linear solutions?  From Figure
\ref{sphcoll}, we learn that virialization is flagged by the
recollapse of the outermost shell.  The value of the linear $\delta$
at the time of collapse depends on the cosmic expansion rate, and
therefore on the cosmological parameters.  Thus, the linear value of
the overdensity can be used as a flag for collapse, providing a simple
and convenient criterion to decide when a perturbation is virialized.

If we assume that the radius of the virialized halo is about half of
the radius of maximum expansion, the reader should be able to derive
the actual average density contrast within the virialized halo with
respect to the ambient density, as well as the linear value of the
density contrast at the time of collapse.  This can be left as a
useful exercise, in the simple case of an EdS universe (see Kaiser
2002 for the solution and much more on cosmic stucture formation).  It
turns out that the linear threshold for collapse in an EdS universe is
$\delta_{c0} = 1.686$, while the actual density contrast of a
virialized halo is $\Delta_{vir} = 178$.  These numbers, particularly
the linear threshold, generalized for different choices of the
cosmological parameters, will be relevant for the following analysis.
One may wonder how few magic numbers can describe a plethora of
complex physical processes.  However, as we will see, these numbers
allow us to make several predictions, whose reliability is supported
by numerical experiments.  We have many reasons to proceed
confidently.

\subsection{Clusters of galaxies reflect the expansion rate of the 
Universe}



As we saw, the expansion rate of the Universe, entering equation
\ref{deltaevol}, affects the evolution of the linear perturbations.
It follows that the growth of a perturbation is slower when the
expansion is faster.  The Hubble parameter in its general form writes
$H(z) = H_0 [ \Omega_0(1+z)^3+\Omega_k (1+z)^2 + \Omega_\Lambda
(1+z)^{3+3w}]^{1/2}$, where $w$ is the ratio between the pressure and
the energy density in the equation of state of the dark energy
component (Caldwell et al. 1998).  The special case $w=p/\rho = const
= -1$ corresponds to the quantuum vacuum energy, {\sl aka} the
cosmological constant.  We find that in a low density Universe the
expansion is faster than in the EdS case, so we expect that clusters
form much later for the same initial conditions.  We have a situation
more similar to a low density Universe in the case of a flat Universe
with cosmological constant.  This last case is the favorite choice,
since today many observational evidences tell us that the Universe is
accelerating (as shown in several lectures at this School), and in the
context of general relativity, this can be explained by the presence
of dark energy.

Therefore, if we set the initial conditions and the cosmological
parameters, we can predict the redshift when virialized halos of a
given mass are expected to form.  At this point we can reverse the
problem: given a measure of the initial conditions (the fluctuations
in the CMB are providing them at a redshift $z\sim 1500$) and after
counting clusters of galaxies at each redshift, we can infer the
expansion rate of the Universe and therefore the cosmological
parameters.  Clusters are much more useful for this kind of test than,
for example, galaxies, since they are the largest virialized structure
in the universe, therefore the closest to the initial linear spectrum
of density perturbations and most affected by the expansion rate.

To play this game, obviously we should not focus on single objects,
rather we should measure the evolution of the number density of
clusters with the cosmic epoch and their distribution with mass.
Let's see this in detail.

\subsection{Where cosmological parameters enter the game}

We are interested in the statistical properties of the initial
conditions, in other words, to the average value of $\delta$ on a
given scale.  Since the majority of inflationary models predict that
the fluctuations in the density field $\rho$ should be Gaussian, we
need to know only its variance.  To define operationally the variance
on a given scale, we can imagine to smooth the linear field by
measuring the overdensity around each point in space within a sphere
of radius $R$ (the top--hat filter function).  Since the
density field is linear, a spatial scale is related to a mass scale
simply by $M = (4\pi /3) \rho_0 R^3$ where $\rho_0$ is the average
density.  If we express the fluctuations field  in terms of
its Fourier power spectrum $P(k)$, the variance reads:

\begin{equation}
\sigma^2(M) = {1\over {8 \pi^3}} \int W^2(kR) P(k) d^3k \, ,
\label{sigma}
\end{equation}

\noindent
where $W(kR)$ is the filter function in the Fourier space.  The filter
function corresponding to a top--hat in real space is oscillating due
to the sharp edges (see Kolb \& Turner 1990, Kaiser 2002).  Since
every mode grows independently from each other in the linear regime,
we expect that $\langle \delta \rangle$ is proportional to the linear
growth factor $D(t)$.  If we call $\delta_{c}$ the critical value
corresponding to the collapse, the epoch of collapse of an overdensity
of a mass $M$ is implicitly defined by the relation:

\begin{equation}
\sigma(M) D(z_{coll}) = \delta_c \, .
\label{cond}
\end{equation}

\noindent
The linear growth factor $D(z)$, which we know since it is the
solution of equation \ref{deltaevol}, can be written in a generic
cosmology as follows (see Peebles 1993):

\begin{equation}
D(z) = {5 \over 2} \Omega_0 E(z) \int^\infty_z {{1+z'}\over{E(z')^3}}
 dz'\label{dz}  \, ,
\end{equation}

\noindent
where $E(z) \equiv H(z)/H_0$.  In the general case it is not possible
to invert analytically equation \ref{cond}.  Again, since we are still
in the theoretical mood, we can assume the EdS case and enjoy its
analytical formulae (as you see, simplicity sometimes attracts
theoreticians against any evidence!).  Another useful step is to
approximate the linear spectrum of the density perturbations with a
power law, $P(k)\propto k^n$ with $n\simeq -1; -2$.  In this case,
from equation \ref{sigma}, $\sigma(M)\propto M^{-a}$ with $a = (n+3)/6
>0$.  Since the linear gowth is $D(t) = (t/t_0)^{2/3} = 1/(1+z)$, we
easily can invert the condition $D(t)\sigma(M) = \delta_{c}$ to obtain
the typical mass which is collapsing at a given epoch:

\begin{equation}
 M_c(t) = M_{c0} (t/t_0)^{4\over {n+3}} \, .
\label{mc}
\end{equation}

\noindent
Here we meet a fundamental property of any model based on CDM: the
hierarchical clustering.  For any $\sigma(M)$ which is decreasing with
mass (which implies $a > 0$, or $n>-3$), more massive objects form at
later times.  The hierarchical clustering, i.e., the progressive
assembling of larger and larger structures with cosmic time, is the
direct consequence of this property.  Actually, the preferred choice
is the spectrum for adiabatic fluctuations in a CDM universe, and it
is the result of a detailed computation involving fluid equations for
relativistic and non--relativistic components in an expanding universe
(see the software CMBFAST, http://cmbfast.org/, by U. Seljak and
M. Zaldarriaga).  Unsurprisingly, a realistic CDM spectrum is not as
simple as a power law.  The resulting $\sigma(M)$ shows a varying
slope as shown in Figure \ref{spectrum}.

\begin{figure}
\centering
\vskip 3.5cm
\begin{center}
{\bf see fig3.gif}
\end{center}
\vskip 3.5cm
\caption{The typical value of the linear fluctuations $\sigma(R)$
predicted for an $\Omega_0=0.3$, $\Omega_\Lambda=0.7$ Universe compared with
the values obtained from observations on different scales (from
Tegmark 2002).  See lectures by W. Percival for constraints from SDSS,
and by R. Caldwell for constraints from the CMB).}
\label{spectrum}     
\end{figure}

Before ending this section, we remark that few years ago, the
hierarchical clustering hypothesis was not so radicated into
cosmological models.  Imagine that we have a kind of dark matter which
has no power at all at small scales.  From equation \ref{sigma}, it is
easy to see that $\sigma(M) = const $ below some threshold $M<M_{th}$.
As a consequence, all the scales with $M<M_{th}$ collapse at the same
time. If $M_{th}$ is large enough, let's say the scale of a cluster of
galaxies, then clusters form at the same time or even before galaxies.
This is the situation we have when we consider light particles like
massive neutrinos as candidates for dark matter.  Now we know that
neutrinos give a negligible contribution to the density of the
Universe (see Lahav \& Liddle 2006).  Given the success of the CDM
spectrum in reproducing observations on several scales (as shown in
Figure \ref{spectrum}) the common wisdom is that cosmic structures
follow hierarchical clustering, at least as far as dark matter is
concerned (but beware of the baryons!\footnote{As a further
complication, there are now strong hints that massive galaxies form
earlier than smaller ones, and bright quasars peaks earlier than
weaker AGN.  This anti--hierarchical behaviour of the stellar mass
component and nuclear activity could, in principle, be reconciled with
the hierachical clustering of dark matter halos.  But this is a
debated issue, known as the ``hierarchical versus monolithical''
controversy.  People use to get very aggressive on this topic.}).

\subsection{The mass function}

Now we can predict the typical mass scale which is virializing at a
given redshift.  Is this enough to efficiently constrain the
cosmological parameters?  Not yet: we can work out much better
observables which will allow us to perform efficient cosmological
tests.  The important step we have to take now is to derive the mass
function.  To do that, first we must write the probability
distribution of the fluctuations $\delta$, which we already assumed to
be a Gaussian with dispersion $\sigma$:

\begin{equation}
P(\delta) = {1\over {\sqrt{2\pi}\sigma}} exp\{{1\over 2}
\delta^2/\sigma^2\} \, .
\end{equation}

\noindent
Since we are dealing with a linear field, it seems safe to say that
the fraction of mass which is in regions with overdensity larger than
a given $\delta$, is equal to the fraction of volume that, filtered
with our top--hat filter of size $R$, is overdense above the same
threshold.  This fraction is simply the integral of the Gaussian from
the overdensity threshold up to infinity.  If we set $\delta_c(z) =
\delta_{c0}\times D(0)/D(z)$, we obtain the fraction of mass which is
in virialized halos at a given epoch $z$.  This condition reads:

\begin{equation}
N(M)M dM  =  \int_{\delta_c(z)}^\infty
P(\delta, \sigma(M)) d\delta  \, ,
\end{equation}

\noindent
where $N(M)$ is the number density of virialized halos in the mass
range $M$ and $M+dM$.  Then we can obtain an expression for $N(M)$
simply deriving the integral on the right hand side with respect to
mass:

\begin{equation}
N(M) = {{\rho}\over {M}} {d\over {dM}} \int_{\delta_c(z)}^\infty
P(\delta, \sigma(M))  \, .
\end{equation}

\noindent
Our tidy theoretical attitude is rewarded again: the solution is
analytic.  Leaving the mathematics to the reader, we write the final,
famous result, the Press \& Schechter (PS) mass function (1974):

\begin{equation}
N(M) \, dM \, = \, \sqrt{{2\over \pi}} {\rho \over M} {{\delta_c(z)}\over
{\sigma^2}} {{d\sigma}\over{dM}} \, exp\Big( -
{{\delta_c(z)^2}\over {2 \sigma^2}}\Big)\,  dM  \, .
\end{equation}

\indent Its typical shape is characterized by a power law at low
masses, and an exponential cutoff at large masses.  Given its
simplicity, its success is often referred as {\sl the Press \&
Schechter miracle}.

\subsection{Is the Press \& Schechter approach accurate enough?}

\begin{figure}
\centering \includegraphics[height=5cm]{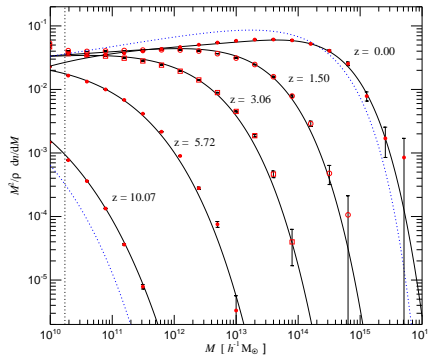}
\caption{The Press \& Schechter mass function (dotted lines) tested
against N--body simulations (dots and solid lines, from Springel et
al.  2005).}
\label{nbody}     
\end{figure}

Unfortunately miracles are not allowed in science.  You may think that
this approach is just a didactical exercise to understand the basic
concepts, while cosmologists actually use terribly complicated
formulae or awfully long numerical computations for the mass function.
Well, the truth is that this formula is still at the core of the
majority of the works deriving cosmological parameters from clusters
of galaxies.  Indeed, many numerical experiments (N--body simulations)
actually support the validity of the PS approach.  Clearly, some
differences with respect to the original PS approach were found.
Discrepancies are mostly due to the many non linear effects which are
not included in the PS formalism.  A recent example of a comparison
between N--body and the PS formula is shown in Figure \ref{nbody}.  We
note that the PS formula tends to underestimate the number density of
halos at very high redshift.  However, if we consider that clusters
are observed today up to redshifts slightly above 1, we have to admire
the remarkable similarity with the results from the time--expensive,
brute--force approach of N--body simulations.

To improve the PS model, some empirical fitting formulae were proposed
on the basis on N--body simulations (Jenkins et al. 2001).  However,
this approach is heavy, because in principle it requires a new
simulation every time the cosmological parameters are varied.  The PS
mass function has the great advantage that the cosmological parameters
space can be explored rapidly.  Finally, I just mention here that the
PS formalism can be extended to give complete merger histories of
single halos (Lacey \& Cole 1993), conditional probability function of
progenitor halos (Bower 1991), biased distribution of halos within
halos (Mo \& White 1996), all topics we do not explore here, but that
proved to be very useful in interpreting data.  As a final comment,
the PS approach after more of 30 years, is still extensively used in
the large majority of the papers on precision cosmology with clusters
of galaxies.

\subsection{From the mass function to the distribution of observables}

Let's take a closer look to the behaviour of the mass function.  We
identify two sets of ingredients: the initial conditions
(normalization and shape of the power spectrum) entering $\sigma(M)$,
and the cosmological parameters ($\Omega_0$, $\Omega_\Lambda$, $w$)
entering $\delta_c(z)$ and the overall normalization of the mass
function.  As you can easily see, the exponential cut off at the
massive end is where the function is most sensible at both sets of
parameters, through the normalization of the power spectrum (expressed
conventionally as $\sigma_8$, which is the amplitude of the spectrum
at the scale of $8 h^{-1}$ Mpc), and the linear growth factor.  We can
now see in much more detailed terms the behaviour we already
appreciated qualitatively: the evolution of cosmic structures is
slower in a universe with lower density with respect to and EdS
universe; the same for a $\Lambda$--dominated, flat universe.  In
quintessence models, for higher values of the parameter $w$, the
growth ceases earlier.  If we normalize our mass function in order to
have the same local density of clusters today, the evolution with $z$
appears faster for $\Omega_0=1$ than for open or $\Lambda$--dominated
universes.  This is shown visually by the N--body simulation in Figure
\ref{evol} (upper panel).  Quantitatively, the expected evolution of
the number density of massive clusters (with virial mass $M> 5 \times
10^{14} h^{-1} M_\odot$) is also shown in Figure \ref{evol} (lower
panel).  Now our observational side can take over, and note that,
after all, an EdS universe is not more appealing than a low density
one since, after all, in the last case we expect much more clusters at
high redshifts.  And observers love to find high--redshift objects.

\begin{figure}
\vskip 5.5cm
\begin{center}
{\bf see fig5.gif}
\end{center}
\vskip 5.5cm
\caption{Top: clusters of galaxies (circles) in an N--body simulation
for an EdS universe (bottom panels) compared with clusters in an open
FRW universe, with statistically equivalent conditions at $z=0$.  The
evolution backward in time of the mass function is strikingly
different (Borgani \& Guzzo 2001).  Bottom: the evolution of the
number density of clusters with virial mass $M> 5 \times 10^{14}
h^{-1} M_\odot$ for different choices of the cosmological parameters
(Rosati, Borgani \& Norman 2002).}
\label{evol}     
\end{figure}

We are almost ready to handle real data, except for a final, small
step, which consists in a simple change of variables.  As you know, in
most cases we do not measure directly the virial mass.  What an
astronomer typically measures is the emitted light in a given band.
In our case, as we will see shortly, we will focus on measuring the
total luminosity $L$ in the X--ray band and the virial temperature $T$
of the diffuse gas.  Therefore, we prefer to have a prediction for the
luminosity or the temperature function.  This is straightforward if we
have a relation $M$--$L$ or $M$--$T$.  We know, from the virial
theorem, that these relations can be obtained from our spherical
collapse model.  Once we have the relationships between the
observables and the mass, we can write the luminosity (XLF) and the
temperature (XTF) functions as:

\begin{equation}
\Phi(L)dL = N(M) {{dM}\over{dL}} dL \, , \, \, \, \, \, \,
\Phi(T) dT=  N(M) {{dM}\over{dT}} dT \, .
\end{equation}

\noindent
Enough theory.

\section{From Observations to Cosmological Parameters}

\subsection{The observer's mood}

We can start the second part of this lecture, where we will encounter
different kind of problems.  We are about to look at data, therefore
we will face reality, which is always somewhat shocking when coming
from the ideal, linear and spherical world of theory.

As we already know, we need a good measure of the actual number
density of clusters of galaxies as a function of mass and redshift.
We also know that we will get the luminosity or, in the best case, the
temperature function of clusters.  This implies that we need to be
able to: find clusters, measure with high accuracy the quantity of
interest, and define the {\sl completeness} of our survey.
Completeness is a key quantity in observational cosmology.  A well
defined completeness means that, for the solid angle of the sky
covered by our survey, we are able to detect all the objects with
luminosity (or temperature, or mass) above a given value and within a
given redshift.  This is mandatory to compute the volume we actually
explore in the survey and, therefore, the comoving number density.
Needless to say, a survey with few objects but a well defined
completeness is way much better than a survey with hundreds of objects
but a poorly defined completeness.  Therefore, we need a strategy to
find as many clusters as possible with a well defined completeness.
Which is the best observational window to do that?  Let's start
examining some options.

\subsection{Optical band}

Searching for clusters in optical images is basically counting
galaxies and looking for overdensities with respect to the background
value (see Gal 2006 for a review).  In doing this, the optical colors
of the member galaxies are a very useful information.  Passive, red
galaxies preferentially populate the central regions of clusters, and
they form a well defined color--magnitude relation.  A galaxy
selection picking the reddest galaxies in the field, helps in reducing
the contamination by the field galaxies.  These techniques can give
efficient clusters detection out to $z\geq 1$ (see Gladders \& Yee
2005).

Optical surveys are very convenient to find many cluster candidates.
We remind that clusters are rare objects (especially the massive ones)
and therefore we need to survey large area to find many of them.  The
optical band offers the opportunity to cover wide area with large CCD
frames, coupled to the availability of ground--based telescopes with
large field of view.  However optical observations have the drawback
of a difficult calibration of the selection function, and therefore
the completeness of an optical survey of clusters is very hard to
define.  This is because the detectability of a cluster depends on the
luminosity, the number, and the concentration of its galaxies, three
aspects that can vary from cluster to cluster.  In addition,
projection effects cause severe contamination from background and
foreground galaxies: filamentary structures and small groups along the
line of sight can mimic a rich cluster.  For the same reason, in the
presence of a positive fluctuations of the background galaxies,
moderately rich cluster can be missed.

More troubles when we try to relate the optical light to the total
mass.  The total optical luminosity of a cluster is somehow
proportional to the total mass.  But we know that the stellar mass in
the galaxies represents a tiny fraction of the total, and usually only
the brightest galaxies are detected, so that a lot of stars in small,
undetected galaxies must be accounted for, by assuming a model for the
galaxy luminosity function.  So, we should not be surprised to know
that the relation between the total optical luminosity of a cluster
and its total mass is very loose.  In order to obtain an accurate
measure of the mass, we may use optical spectroscopy to measure the
velocity dispersion of the galaxies and then apply the virial theorem.
However, this requires a lot of observing time, and obviously it is
still affected by contamination from interlopers.  All these problems
become more severe at high redshift, as the field galaxy population
overwhelms galaxy overdensities associated with clusters.  A
completely different technique is to measure the mass directly through
strong and weak lensing (see, e.g., Dahle et al. 2003).  This is a
very promising tool, but it has its own problems, like severe
projections effects (the lensing depends on all the mass along the
line of sight towards the clusters and on its position) and the
difficulty to obtain clean lensing signal.

\subsection{Millimetric band (SZ effect)}

Among the many virtues of clusters of galaxies, there is this peculiar
feature: clusters can be seen as shadows on the cosmic background
radiation.  This is due to the Sunyaev--Zeldovich (SZ) effect (Sunyaev
\& Zeldovich 1972).  We know that most of the baryons in clusters are
in the form of very hot, ionized gas.  Photons from the CMB passing
through a cluster find many high--speed electrons and therefore
experience Inverse--Compton scattering.  In this process, the energy
is transferred from the electrons to the much colder CMB photons.
Since this process preserves the number of photons, the net result is
that the black--body spectrum of the CMB is slightly distorted and
shifted to larger frequencies by an amount that depends on the
temperature, and on the column density of the ICM.  The net effect on
the CMB is the production of a cold spot at low and a hot spot at high
frequencies, where the pivotal frequency is about 217 GHz (see Holder
\& Carlstrom 2001).  This sounds very promising, since we have both a
spatial and a spectral signature.  Actually, several clusters have
been imaged with the OVRO and BIMA arrays (Carlstrom et al. 2002).
Indeed, the scientific community is making a strong effort to build
instruments that can study both CMB and the SZ effect from the ground
(like AMI, ACT, AMiBA, APEX, SPT), or from space (like the Planck
satellite, whose full--sky survey is expected to detect thousands of
clusters).

Among the positive aspects of SZ observations, we find the absence of
the redshift dimming, which allows one to identify clusters virtually
at any redshift.  This means that the selection criteria are
essentially equivalent to a completeness in mass, which is very
desirable.  However, severe contamination from foreground and
background radio sources is expected.  Multi--frequency observations
can help a lot in disentangling the spectral signature of the SZ
effect from the spectrum of radio sources.  However, the difficulties
in detecting clusters via the SZ effect are still significant (see
Birkinshaw \& Lancaster 2004).  An easy prediction is that in five
years, the SZ effect will be one of the main observational window to
find and study clusters of galaxies.

\subsection{X--ray band}

At present, in my view, the X--ray band is the most convenient  to
find and investigate clusters.  Anyway, it is the field in which I
spent most of my activity, and therefore, for a mix of objective and
private reasons, since now on, I will focus mostly on X--ray.

The first thing to say is that clusters appear as strong--contrast
sources in the X--ray sky up to high redshifts, thanks to the
dependence of the X--ray emission on the square of the gas density
(see \S 2.5).  Given the relatively small number of sources, X--ray
images of clusters are virtually free from contamination from
foreground and background structures.  In other words, clusters are
the second most prominent sources in the X--ray sky (after Active
Galactic Nuclei), at striking difference with the optical and
millimetric bands where they have to struggle to emerge above other
stronger signals.  This can be clearly appreciated in Figure
\ref{ms1137} (left) where almost all the point sources in the image
are AGN, while the bright, extended source in the center is a cluster
at $z=0.79$.  The image has been taken with the ACIS--I detector,
covering a square of 16 arcmin by side.  X--ray emission from clusters
can be detected up to redshift larger than one, as shown in Figure
\ref{ms1137} (right) where the X--ray emission (red) from the
$z=1.235$ cluster RXJ1252 is shown on top of the optical image.  

\begin{figure}
%
\vskip 2cm
\begin{center}
{\bf see fig6.gif}
\end{center}
\vskip 2cm
\caption{Left: The cluster MS1137, $z=0.79$, in a field observed for
116 ks with the X--ray telescope {\sl Chandra}.  The cluster is the
bright extended source in the center, while most of the remaining
sources are AGN.  Right: the X--ray emission from the $z=1.235$
cluster RXJ1252 is shown on top of the optical image taken with the
VLT telescope (Rosati et al. 2004).}
\label{ms1137}
\end{figure}

A flux--limited X--ray survey can provide a sample of clusters with a
well defined completeness, thanks to the fact that the X--ray emission
from clusters is continuous (at variance with the optical emission
associated to the single galaxies) and centrally peaked towards the
center.  Therefore we just need to establish a robust connection
between the X--ray luminosity and the total mass.  A potential problem
with X--ray clusters is that the X--ray flux is sensitive to
irregularities in the gas distribution.  However, this problem does
not seem dramatic, given that most of the clusters appear smooth and
round, and the theory provides us with a robust connection between the
total mass and the ICM properties.  Thus, for the moment, we just need
to fully appreciate the advantages in looking at clusters with X--ray
satellites, which became possible since the 60s thanks to the first
X--ray missions leaded by Riccardo Giacconi.  In the spirit of
constraining the cosmological parameters, X--ray surveys of clusters
of galaxies had a large success in the 90s, thanks to ROSAT and other
satellites, and provided consistent but sometimes debatable results.
For a review of the many surveys with cosmological impact see the
review by Rosati, Borgani \& Norman (2002).

We are now in the era of the XMM--Newton and Chandra satellites.
These two telescopes are mostly performing pointed observations of
clusters discovered in the previous surveys.  No wide area surveys are
currently planned, given the small field of view of these satellites
(nonetheless, some serendipitous surveys are underway with both of
them).  These pointed observations are bringing to us many beautiful
images, along with many unconfortable news that we will discuss in \S
3.  Before stepping further, let's remind the basics of X--ray
emission from the ICM.

\subsection{The X--ray emission from Clusters of Galaxies}

We know that most of the baryons in clusters are in the form of hot
plasma.  This plasma is optically thin and it radiates by free--free
(bremsstrahlung) emission.  It is in collisional equilibrium,
therefore its typical temperature is set by the large dynamical masses
of clusters ($10^{14}-10^{15} M_\odot$) to be in the range of 10--100
millions K (corresponding to 1--10 keV).  This implies that most of
the emission is in the X--ray band.  The total X--ray emissivity due
to thermal bremsstrahlung is obtained by integrating over the
distribution of speeds of the plasma electrons, and, after a further
integration over frequencies, it can be written as (see Ribicky \&
Lightman 1979):

\begin{equation}
{{{dL} \over {dV}} ~ = ~ 1.4 \times 10^{-27} \, T^{1/2}\, n_e^2 \, Z^2
\, \bar g_B ~ {\tt erg \, s^{-1} cm^{-3}} } \, ,
\label{bremss}
\end{equation}

\noindent
where $Z$ is the atomic number of the ions and $\bar g_B$ is the
velocity--averaged Gaunt factor averaged over frequencies.  First, we
notice the dependence of the total emissivity on the square of the
electron density.  This is the main reason why clusters are
high--contrast sources in the X--ray sky, and also why superposition
or confusion effects due to smaller background or foreground halos,
are less important than in the optical band, where the total
luminosity scales linearly with the (stellar) mass.  We also note the
weaker dependence on the temperature ($T^{1/2}$).

\begin{figure}
\centering \includegraphics[height=6cm,angle=-90]{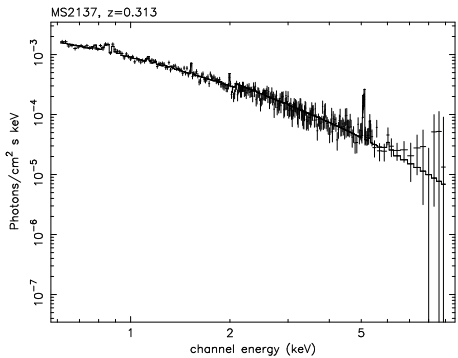}
%
\caption{The spectrum of MS2137 observed with ACIS-I onboard of the
{\sl Chandra} satellite.  MS2137 is a bright X--ray cluster at
z=0.313, with an average temperature of about 5 keV.}
\label{xspec}
\end{figure}

Another contribution to the X--ray luminosity comes from the line
emission due to heavy ions.  This contribution is generally negligible
in terms of total emission, since at temperatures larger than 5 keV,
almost all the heavy nuclei are fully ionized.  However, the
line--emission contribution is increasing at low temperatures, and
starts to be relevant below 2 keV.  This aspect is important when
studying the production of metals in cluster galaxies and their
diffusion into the ICM.  A typical X--ray spectrum of a cluster, with
the typical Iron line at 6.7 keV rest--frame, is shown in Figure
\ref{xspec} (right).

Equation \ref{bremss} gives the luminosity per unit volume, therefore,
the total luminosity must be obtained by integrating up to the virial
radius.  In the simplest assumption of isothermality ($kT = const$ at
any radius in the cluster), the only relevant quantity is the square
of the electron density $n_e^2(r)$, which is generally assumed
proportional to the gas density $n_g$.  In general the gas density
profile is described with the so--called $\beta$--model (Cavaliere \&
Fusco Femiano 1976), which consists in a flat central core and a steep
decrease in the outer regions:

\begin{equation}
n_g \propto 1/(1+(r/r_c)^2)^{3 \beta /2} \, ,
\label{beta}
\end{equation}

\noindent
where $r_c$ is the core radius, and the parameter $\beta \sim
0.5$--$1$ can be interpreted as the ratio of the specific energy of
the dark matter particles (often measured through the galaxies
velocity dispersion) over the gas temperature.  Given the steep slope
outside the core, and the $n_g^2$ dependence of the luminosity, only
the central regions (few core radii) are clearly detected in the
X--ray images.  The outer regions are hardly detected even with
present--day satellites.  Observers often prefer to quote all the
quantities within the observed radius, which is typically half or less
than the virial one.

As we know, X--ray detectors onboard of the Chandra and XMM satellites
are CCD cameras, which read the collected photons every few seconds,
recording both the position and the energy (with a reasonable error of
few percent).  Therefore X--ray astronomy has the big advantage of
recording images and spectra at the same time.  High resolution
X--ray spectroscopy is still feasible through gratings, however the
energy resolution of the CCD is good enough to our purposes of
measuring the temperature of the baryons.

Once we obtain the baryon density from the X--ray surface brightness,
and the temperature of the gas, we can measure the total mass
simply by applying the condition of hydrostatic equilibrium:

\begin{equation}
M(<r) = -{{k_B T R}\over{G\mu m_p}} \Big( {{d{\tt
log}(\rho_g)}\over{d{\tt log}(r)}} + {{d{\tt log}(T)}\over{d{\tt
log}(r)}} \Big) \, ,
\end{equation}

\noindent
where $\mu$ is the mean molecular weight ($\mu \sim 0.6$) and $m_P$ is
the proton mass (see Rosati, Borgani \& Norman 2002).  Here we let the
temperature free to change with the radius.  Of course this equation
is particularly simple in the isothermal case.  In general, the masses
obtained in this way are pretty close to that obtained simply through
the virial theorem $T \propto M^{2/3}$.  On the other hand, it is well
known that clusters do have a temperature structure, which is often
well described by a mild decrease outwards (see Vikhlinin et
al. 2005), and, in more than half of the local lcusters, a drop of
about a factor of three in the very inner regions (the cold core, see
Peterson \& Fabian 2005). The temperature profile is quite important,
but its measure is increasingly difficult at increasingly high
redshifts.  Indeed, we need a lot of photons in order to measure the
temperature in several concentric regions (at least one thousand for
each independent spectrum), and to obtain the deprojected temperature
profiles.  For this reason, virial masses of distant clusters are
often derived assuming isothermality.

Our framework allows us to relate the basic X--ray observables,
luminosity and temperature, to the dynamical mass.  We already know
that luminosity is more affected by the details of the gas
distribution, while the $M$--$T$ relation appears more stable since it
is directly based on the virial theorem.  But we also know that
luminosity is much easier to measure, since we need much less photons
to measure a luminosity, and therefore we can observe many more
clusters within a given amount of telescope time.  A shortcut is to
build phenomenologically the $L$--$T$ relation, fitting the data with
a formula of the kind:

\begin{equation}
L_{bol} = L_6 \Big( {{T_X}\over { 6 keV}}\Big)^\alpha (1+z)^A \Big(
{{d_L(z)}\over{d_{L,EdS}(z)}}\Big)^2 10^{44} h^{-2} erg s^{-1}\, ,
\label{ltrel}
\end{equation}

\noindent
where $\alpha$ is measured to be about 3, while the evolutionary
parameter $A$ is more uncertain and varies between 1 and 0 (see
Vikhlinin, et al. 2002, Ettori et al. 2004).  Once the relations
between the X-ray observables and the mass are established, we can
compare the observed XLF and XTF to our predictions.  For a review of
the X--ray properties of X--ray clusters, see the book by Sarazin
(1988). 

\subsection{Measuring $\Omega_0$ from the observed X--ray luminosity  
 function}

The luminosity function seems easy to measure: first we count all the
clusters in our survey, then we measure their flux just counting the
photons from each cluster.  We also have to know the redshift of each
cluster with a good approximation, in order to compute luminosities.
The redshift can be obtained with an optical spectroscopic follow--up
on a limited number of member galaxies, or with photometric
techniques.  As noted before, shallow X--ray surveys allows us to
measure the luminosity with good accuracy, and to scan a wide area of
the sky. Once we have a flux limited sample with measured
luminosities, we build the XLF by adding the contribution to the space
density of each cluster in a given luminosity bin $\Delta L$ :

\begin{equation}
\phi(L_X) = {1\over L_X} \Sigma^n_{i=1} {1 \over {V_{max}(L_i,f_{lim})}}
\label{xlf}
\end{equation}

\noindent
where $V_{max}$ is the total search volume defined as:

\begin{equation}
V_{max}  = \int_0^{z_{max}} S[f(L,z)] \Big( {{d_L(z)}\over {(1+z)}}\Big)^2
{{c \, dz}\over {H(z)}} \, ,
\end{equation}

\noindent
where $S(f)$ is the sky coverage, which depends on the flux (since the
sensitivity of a survey can vary across the surveyed region of the
sky), and $d_L(z)$ is the luminosity distance.  

\begin{figure}
\centering \includegraphics[height=6cm]{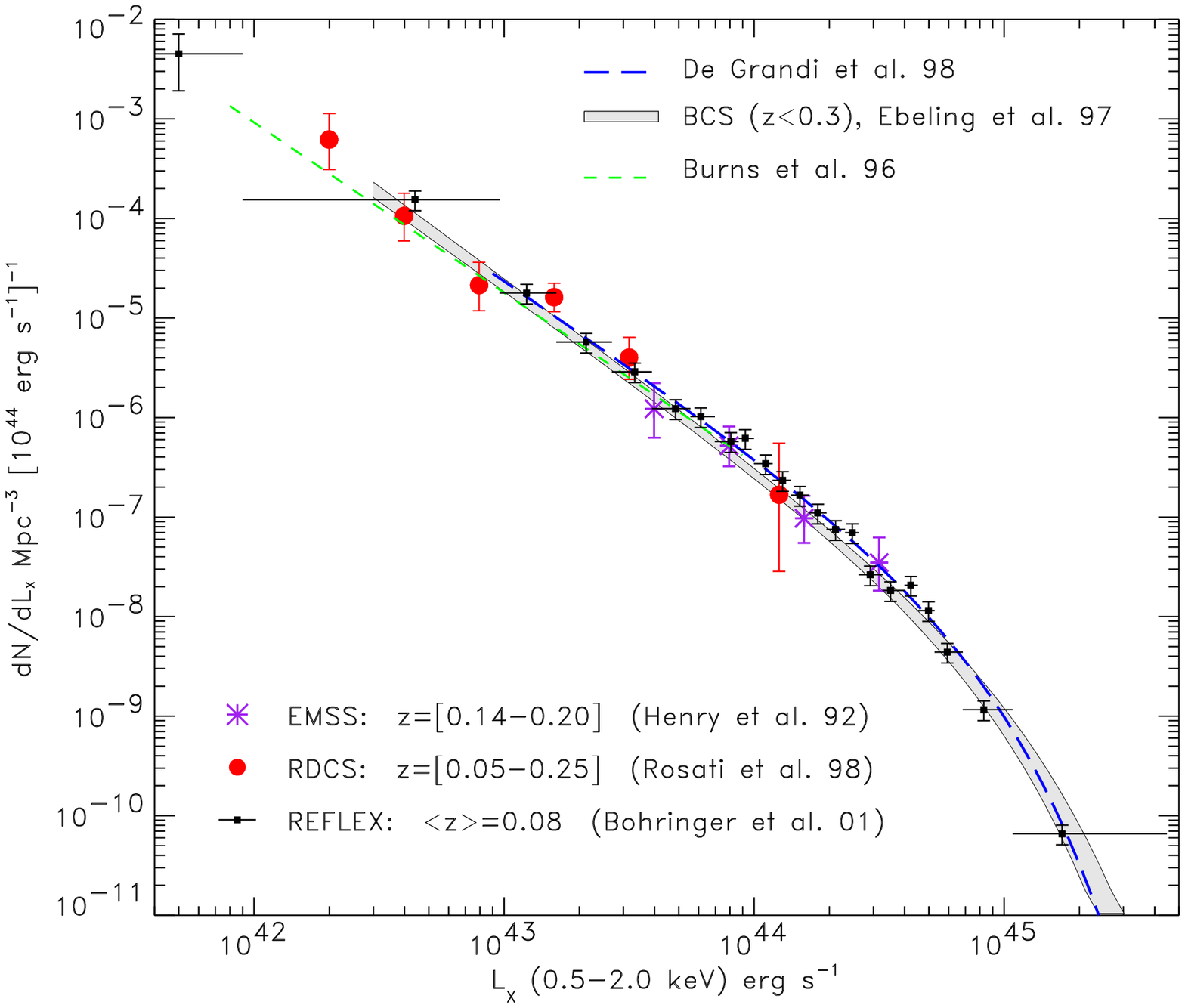}
\centering \includegraphics[height=6cm]{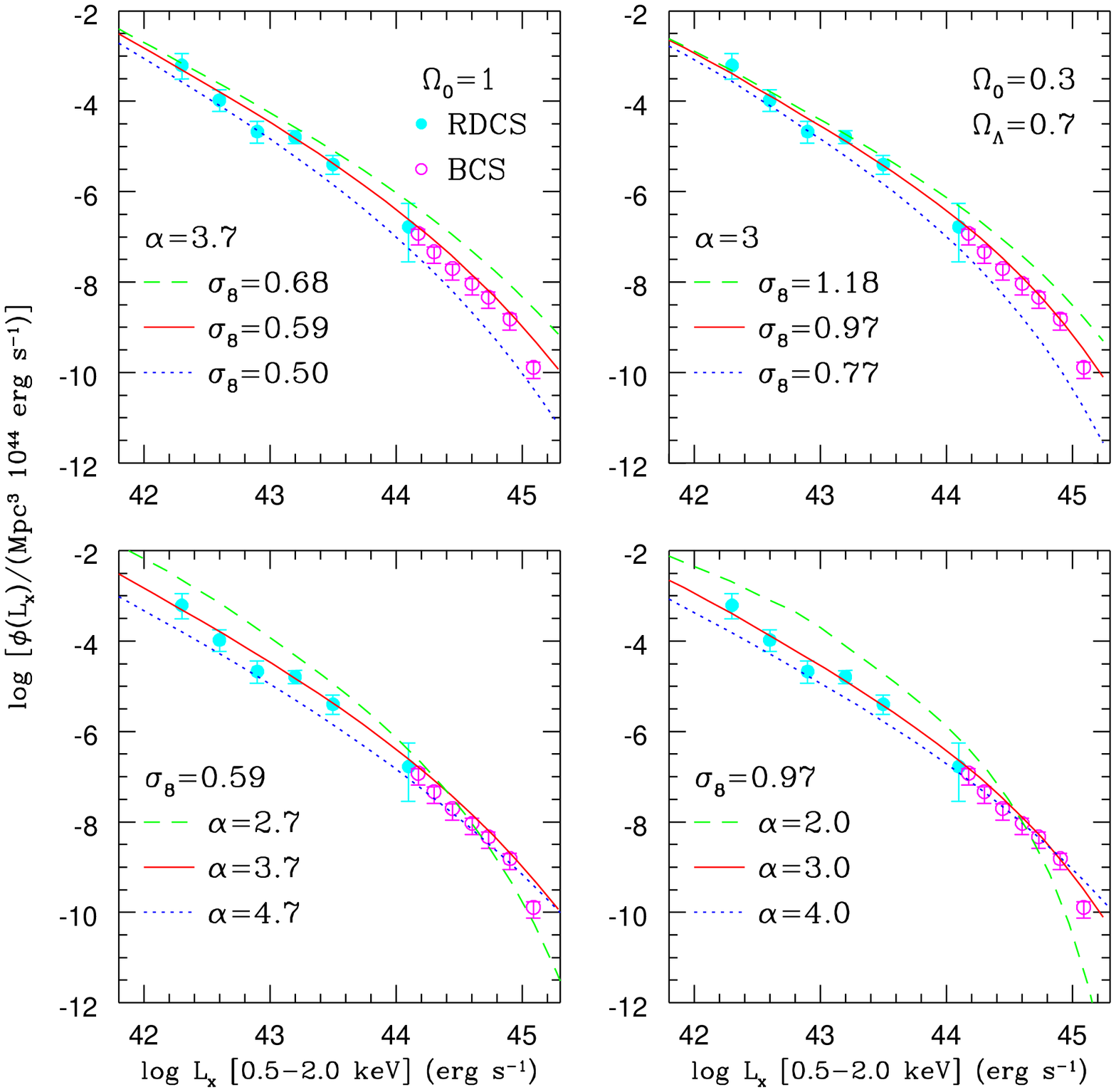}
\caption{Upper panel: the local X--ray luminosity function of clusters
of galaxies from different samples computed for an EdS Universe with
$H_0 = 50$ km s$^{-1}$ Mpc$^{-1}$ (Rosati, Borgani \& Norman
2002). Lower panels: the local X--ray luminosity function of clusters
of galaxies from RDCS (filled circles) and BCS (open circles) for
different $\sigma_8$ and different parameter $\alpha$ for the slope of
the $L$--$T$ relation (Borgani et al. 1999)}
\label{xlf0}
\end{figure}

Remember that we expect to get information on the cosmological
parameters both from the shape of the XLF and from its evolution with
redshift.  To begin with, the shape of the local XLF is well
understood thanks to several different surveys giving consistent
values, and it is shown in figure \ref{xlf0} (upper panel).  This
allows already to get some information from the data at $z=0$, by
finding the parameters which minimize the $\chi^2$ computed on the
binned luminosity function from equation \ref{xlf}, or by a
maximum--likelyhood approach using the unbinned data (see Borgani et
al. 2001).

However, when only local data are used, we find a lot of degenaracy
among cosmological parameters.  Lower $\Omega_0$ can be compensated by
higher spectrum normalization $\sigma_8$ (see Figure \ref{xlf0}, lower
panels).  To break this degeneracy we can use the evolution with
redshift.  The evolution of the XLF is still debated: there is a hint
of evolution at the very bright end, but for the typical $L_*$
clusters and less luminous ones, there is no evolution almost up to
$z\sim 1$ (see discussion in the review by Rosati, Borgani \& Norman
2002).  In other words, most of the clusters, if we exclude the
brightest ones, are already in place at high redshift.  We know what
does it mean, at least qualitatively: the matter density parameter
$\Omega_0$ is significantly lower than 1.

\begin{figure}
\centering \includegraphics[height=6cm]{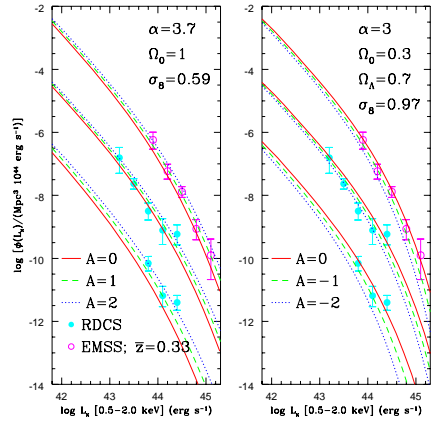}
%
\caption{The X--ray luminosity function of clusters of galaxies
in three different redshift bins: z=0.3--0.6 (EMSS data);
z=0.25--0.50 and z=0.50--0.85 (RDCS).  For each model and at each
redshift, different curves refer to different evolutions for the
$L$--$T$ relation (Borgani et al. 1999). }
\label{b992}
\end{figure}

Our group, few years ago, applied this cosmological test to the RDCS
survey (Rosati et al. 1998), which is the deepest sample of X--ray
selected clusters.  This choice provide a good leverage in terms of
cosmic epoch, but necessarily, given the relatively small solid angle
surveyed with respect to shallower surveys, does not probe well the
high luminosity end.  The results, published by Borgani et al. (1999;
2001) are shown in Figure \ref{b992}, where we used also data from the
EMSS survey (Gioia et al. 1990).  In these Figures we notice that some
degeneracy is still present also when fitting the XLF in the high
redshift bins.  We also notice that the constraints on the
cosmological parameters $\Omega_0$ and $\sigma_8$, are weakened when
the parameters $\alpha$ and $A$, describing the slope and evolution of
the $L$--$T$ relation, are allowed to vary within the observational
uncertainties.

\begin{figure}
\centering \includegraphics[height=4cm]{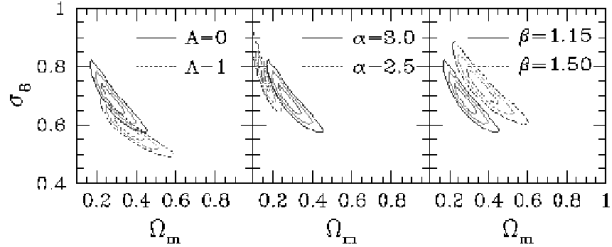}
%
\caption{Confidence contours in the density parameter $\Omega_0$ and
the normalization of the density fluctuations spectrum $\sigma_8$ from
the fit of the high--z XLF for different choices of the parameters
describing the physical relations $L$--$T$ ($\alpha$ and $A$) and
$M$--$T$ ($\beta$; Borgani et al. 2001). }
\label{b01}
\end{figure}

The uncertainties on the cosmological parameters are better shown in
terms of confidence contour levels, where we can also evalutate the
effects of the uncertainties associated to the parameters describing
the $L$--$T$ relation.
5
In Figure \ref{b01} we show how the confidence contours in the
$\Omega_0$--$\sigma_8$ space dance around when the slope and evolution
of the $L$--$T$ relation (parametrized by $\alpha$ and $A$ like in
equation \ref{ltrel}), but also the normalization of the $M-T$
relation (parameter $\beta$), are allowed to vary.  The displacements
of the contours are at more than 3 $\sigma$, therefore we are learning
unconfortable news: the uncertainties on the properties of the ICM are
affecting the cosmological tests at a significant level.

\begin{figure}
\centering \includegraphics[height=4cm]{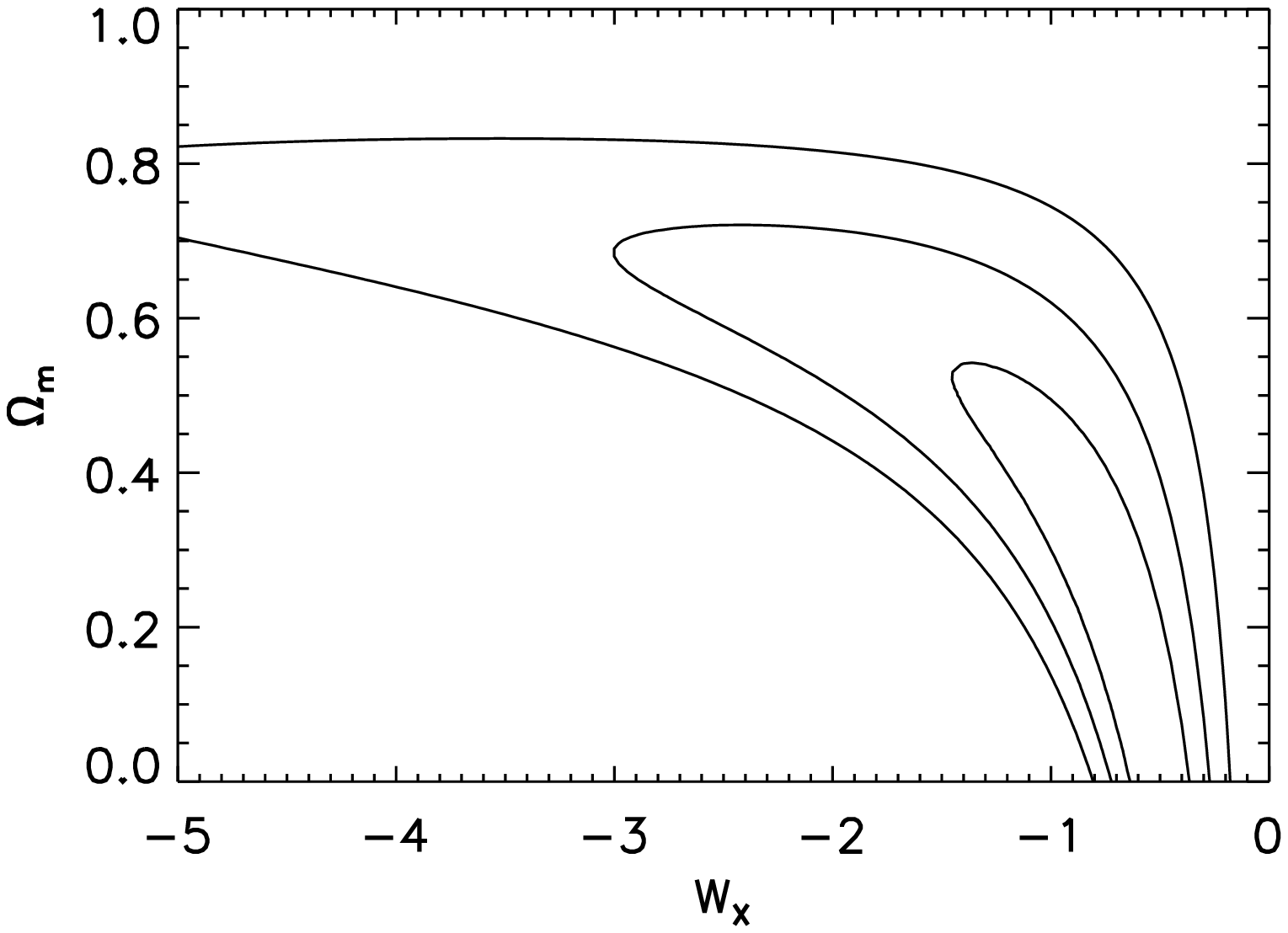}
\centering \includegraphics[height=4cm]{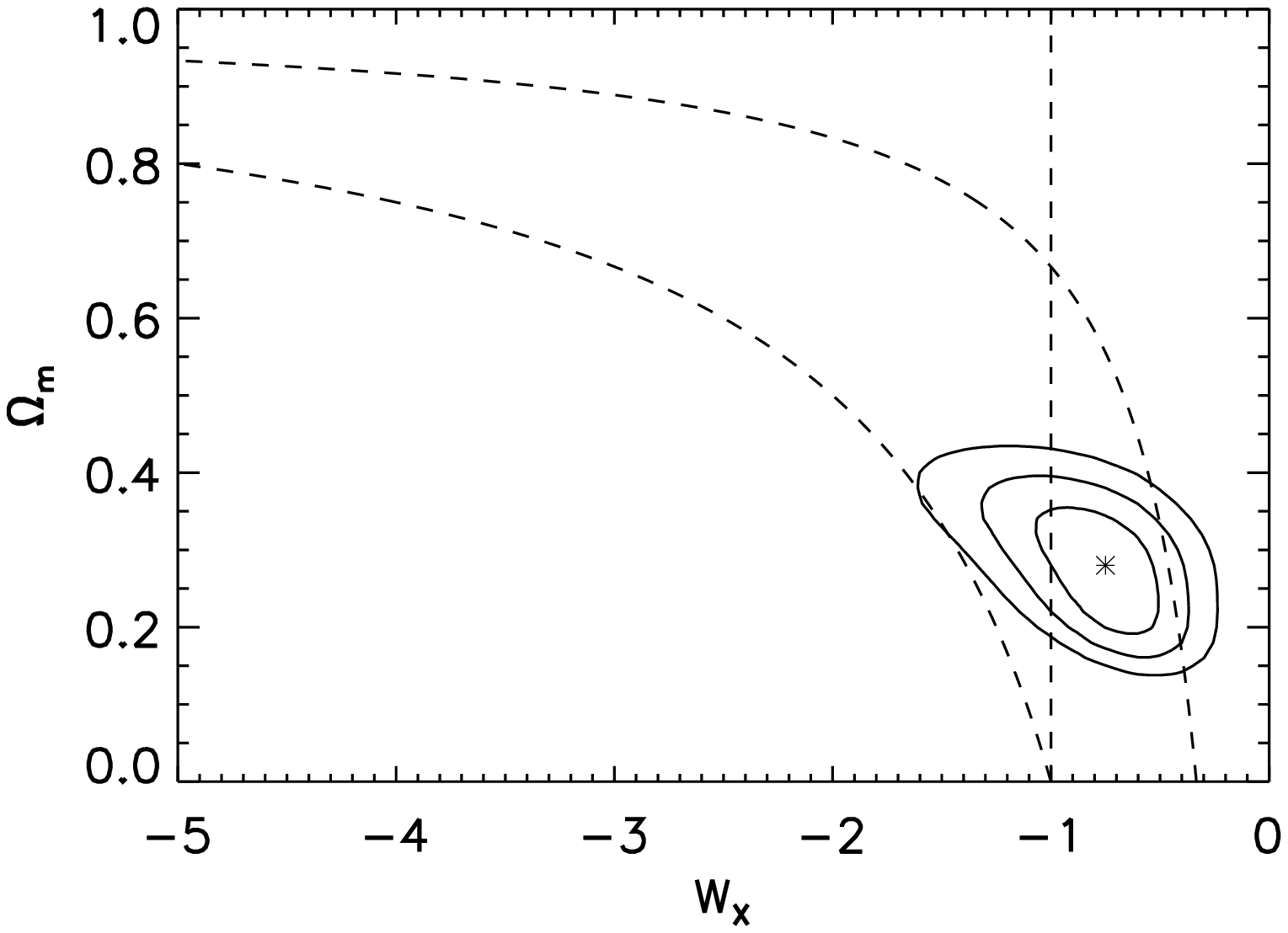}
%
\caption{Confidence contours (1--3 $\sigma$ levels for two degrees of
freedom) in the $\Omega_0$--$w$ plane obtained from SNeIa only
(Perlmutter et al. 1999 sample, left panel) and SNeIa plus REFLEX
(right panel, from Schuecker et al. 2003).}
\label{schu_f4}
\end{figure}

The situation is getting worse when we investigate the dark energy
parameter $w$.  While the density parameter $\Omega_0$ is well
constrained by clusters, $w$ is hardly constrained at all.  Recent
works trying to constrain dark energy, combine constraints from both
SneIa and clusters, to significantly improve the constraints on $w$
due to the complementarity of the two tests in the $\Omega_0$--$w$
space (see Figure \ref{schu_f4}).

%

\subsection{Measuring $\Omega_0$ from the observed X--ray 
temperature function}

At this point you may ask: since our theoretical framework seems
quite successful, why do we have such large uncertainties in the
relations between $L$ and $T$?  Not only we showed that the relation
between mass and luminosity is reasonably understood on the basis of
the spherical collapse, but we also mentioned a possible shortcut
through the direct measure of the $L$--$T$ relation.  Well, we knew
that something wrong were lurking somewhere... However, before
worrying too much, let's give a try to the XTF, which is based only on
the more robust $M$--$T$ relation.  Indeed, the $M$--$T$ relation
relies directly on the virial theorem and it is observed to have
smaller scatter with respect to that observed in the $L$--$T$
relation.

\begin{figure}
\centering \includegraphics[height=6cm]{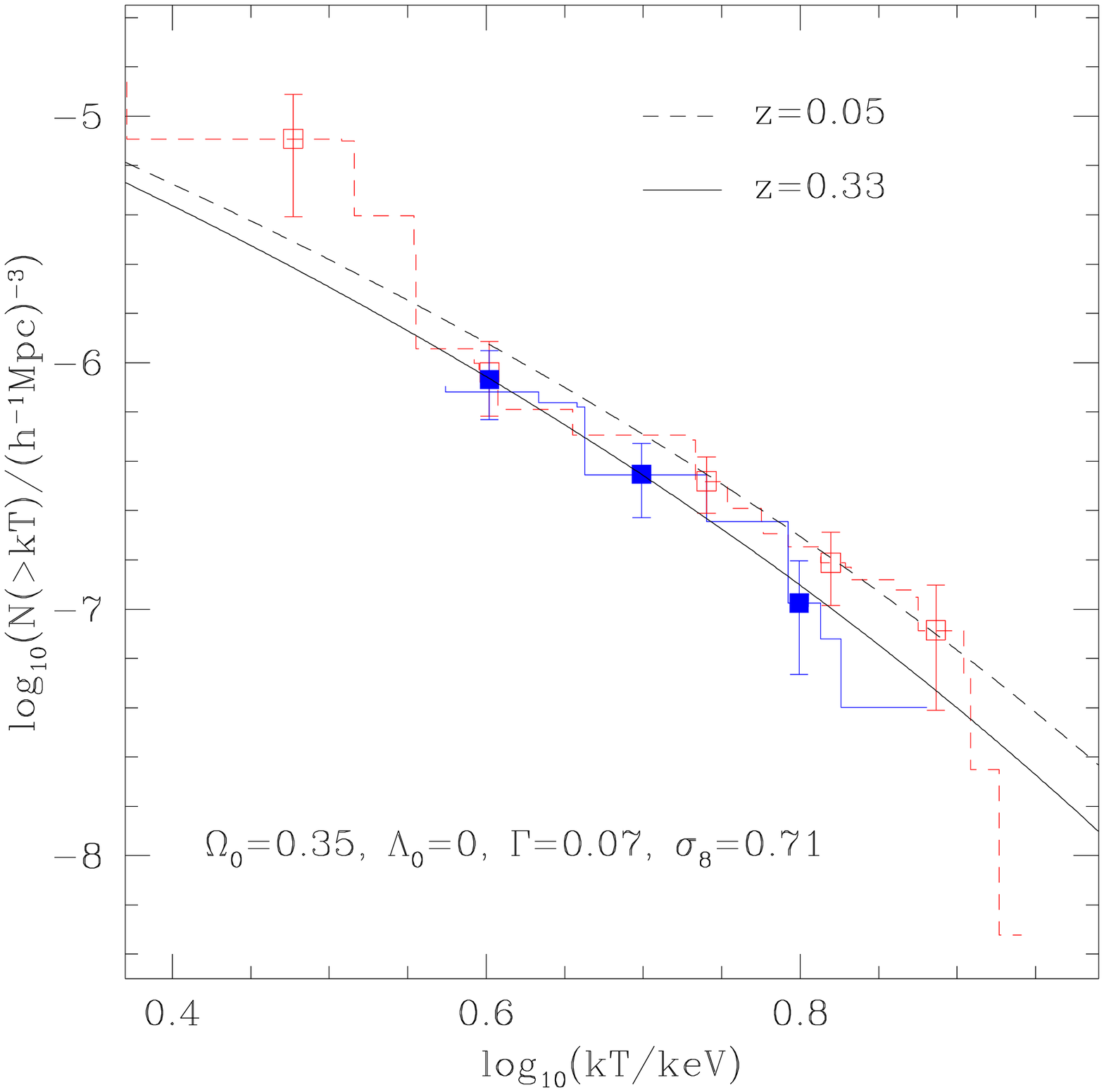}
\centering \includegraphics[height=6cm]{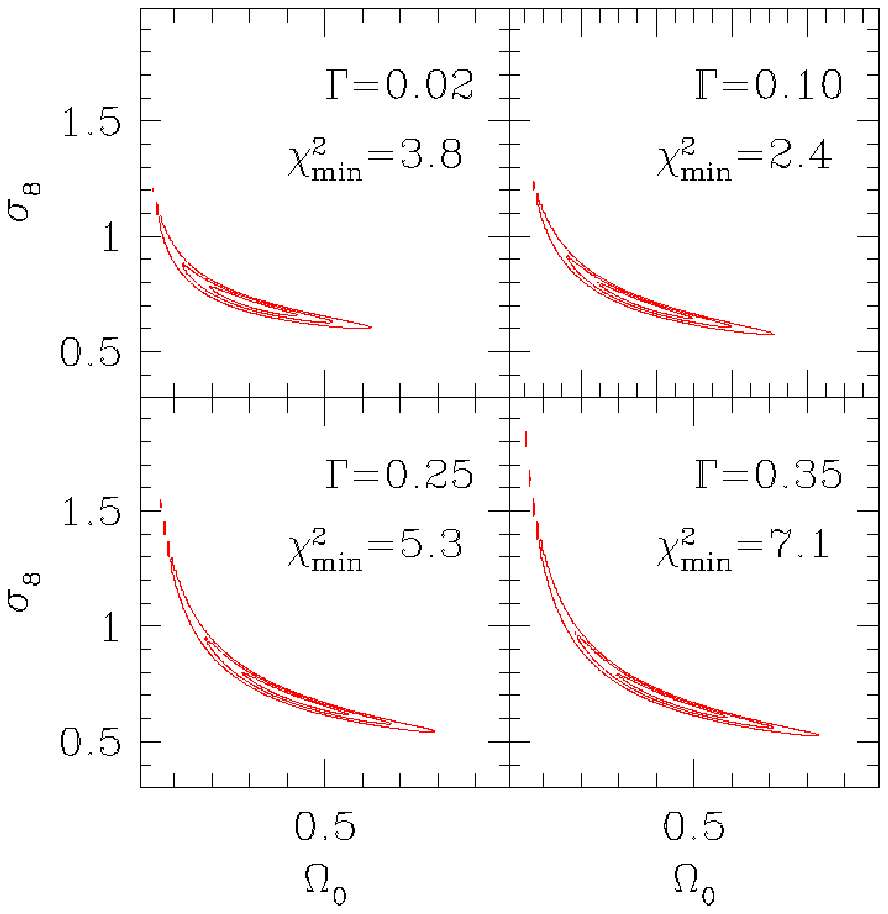}
\caption{Left: Fit to the evolved temperature function. Right:
confidence contours in the $\Omega_0$--$\sigma_8$ space (from Eke et
al. 1998). }
\label{eke98}
\end{figure}

When using the XTF, the price to pay, as we know, is that it is much
more difficult to assemble a complete sample of clusters with
temperatures measured with reasonable errors.  However the XTF is
considered to be more effective in constraining cosmological
parameters.  The first good news is that the constraints from the XTF
are similar to that from the XLF.  The constraints obtained from the
XTF point towards $\Omega_0 \sim 0.3$ for a flat universe (see Donahue
\& Voit 1999), providing at the same time significant constraints on
the normalization of the power spectrum (see Pierpaoli et al. 2001).
In Figure \ref{eke98} we show the results from Eke et al. (1998).  We
notice the tight constraints, but, again, also a significant
degeneration in the $\sigma_8$ --$\Omega_0$ space.

An additional problem comes from a parameter which we considered, so
far, pretty robust: the normalization of the $M$--$T$ relation.  It
has been noticed that the value of $\beta$ found in N--body
simulations is higher than the observed one.  This can be due to
several effects (see Borgani et al. 2004), but the net result is that
the uncertainties on this parameter introduce uncertainties in the
constraints from the XTF in the same way as the $L$--$T$ parameters
are weakening the constraints from the XLF (see, e.g., Ikebe et
al. 2002).


It is clear at this point that the main uncertainties comes from the
poor understanding of the scaling relations between the ICM
observables and the mass, both from the theoretical and the
observational points of view.  A detailed investigation of the effects
of such uncertainties is given in Pierpaoli et al. (2003). They
conclude that the cosmological constraints from XLF and XTF, both the
local and the evolved ones, are reliable and consistent with each
other, but that the statistical errors on the cosmological parameters
are larger than previously thought.  The buzzword now is: we need to
improve the quality of the data on single clusters to better
understand the physics of the ICM.  But why did clusters prove to be
such a difficult topic, after being the best candidate for the most
friendly objects in the Universe?

\section{New Physics and future prospects}

\subsection{Something is missing: new physics for the 
clusters baryons}

Why do we have such a poor understanding of the $L$--$T$ relation?
From equation \ref{bremss}, assuming $n_e \propto \rho_{tot}$ (in
other words, that the baryons follow the total matter distribution),
and integrating over the volume, we obtain $L \propto T^2$ (without
including line--emission).  This is the $L$--$T$ relation predicted in
what is called the {\sl self--similar} scaling (Kaiser 1986).  As long
as the baryons are distributed in the same way of the total mass, each
X--ray observable scales like some power of the mass.  Another way to
say this, is that small clusters are the mass--rescaled version of
massive clusters.

So far, we reasonably expected that the thermodynamics of the ICM,
being dominated by dark matter, is driven by gravitational processes,
like shocks and adiabatic compression occuring during the
virialization phase and the subsequent growth in mass by accretion.
This self--similar behaviour is also supported by N--body
hydrodynamical simulations which do not include radiative cooling.
But the observed slope of the $L$--$T$ relation is much steeper then
predicted ($\alpha \geq 3$ rather than 2 or lower when line emission
is included) and it constitutes the first strong evidence of something
wrong in the self--similar picture.  That's why when performing the
cosmological tests, we avoided this inconsistency by varying the
parameters of the ICM scaling relations.  

\begin{figure}
\centering \includegraphics[height=9cm]{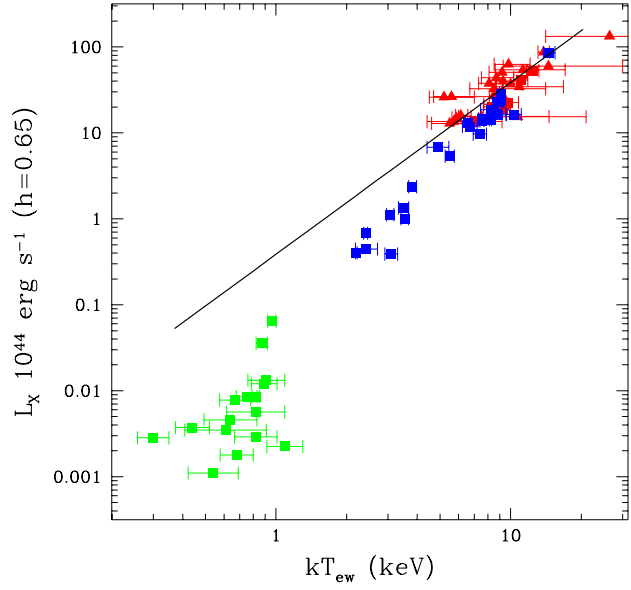}
\centering \includegraphics[height=10cm]{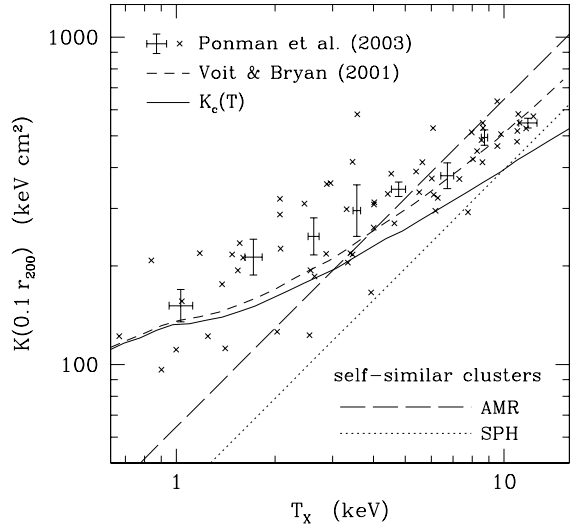}
\caption{Top: the $L$--$T$ relation for groups and clusters showing
the steeper slope with respect to the self--similar model $L\propto
T^2$ (continuous line).  Bottom: the entropy ramp, showing the higher
entropy in low temperature systems with respect to the self--similar
model (Voit, Kay \& Bryan 2005).}
\label{lt}
\end{figure}

However, we learned that thawing the thermodynamic parameters
introduces large uncertainties in the cosmological constraints.
Obviously, we would appreciate a lot to have a physical basis for the
observed scaling relations, in order to better control the
uncertainties due to a poor description of the ICM thermodynamics.
The first step is to invoke a physical process that leads naturally to
an $L \propto T^3$ scaling, in other words, a process which implies a
progressive decrease of the X--ray luminosity at low mass or
temperatures, as shown in Figure \ref{lt} (top).  How can we obtain
this?  We know that we can efficiently decrease the predicted
luminosity by imposing a lower density in the central regions of the
clusters. To do that, we simply need to add an extra pressure, or some
extra amount of energy in the center of clusters.  {\sl Extra} means
in excess with respect to the energy acquired through shocks and
adiabatic heating.  This extra energy does not translate in an higher
temperature; what happens, is that the pressure increases, and the gas
distribution gets puffier, readjusting itself in the dark matter
potential well.  A useful quantity to describe such behaviour is $K
\equiv T/n^{2/3}$.  This is the normalization of the equation of state
of the ICM, which is that of a perfect gas, $p = K \rho^{5/3}$.  We
remind that the entropy is $ S = N ln(K)$.  The entropy is also a very
convenient thermodynamic variable, since it is constant during
adiabatic compression, and it changes only in the presence of
radiative cooling or shock heating.  For this reason, another way of
describing the break of the self--similarity in clusters, as shown by
Ponman et al. (1999), is to plot the entropy as a function of the
cluster temperature, as shown in Figure \ref{lt} (bottom).

The desired effect is obtained by giving about half or 1 keV to each
gas particle.  The effect is small in rich clusters, where the virial
temperature is around 10 keV, and the energy budget is largely
dominated by gravity, while it is increasingly large at lower
temperatures, when the extra energy starts to be a significant
fraction of the gravitational energy scale.  In this way we solved the
problem from the point of view of the thermodynamics (see, e.g., Tozzi
\& Norman 2001).  Of course, the real problem starts now: which is the
physical mechanism responsible for the energy (or the entropy) excess?

We have two obvious candidates which can inject energy associated to
non--gravitational processes: the prime candidate is feedback from
star formation processes, whose effects are testified by the presence
of heavy elements in the ICM.  The second candidate is feedback from
nuclear activity in the clusters galaxies.  Actually, the interaction
of AGN jets and the ICM has been directly observed.  The most
spectacular example is the Perseus cluster, where jets from the
central AGN (visible in the radio emission) is pushing the ICM
creating two large simmetric cavities towards the center (Fabian et
al. 2002; 2003; see Figure \ref{bullet}).  Chandra and XMM added other
surprises: the presence of {\sl cold fronts} (Vikhlinin et al. 2001,
Mazzotta et al. 2001) and of massive mergers strongly affecting the
dynamical equilibrium.  To this, we must add the puzzling discovery by
XMM that the ICM in the central regions never cools down by more than
a factor of 3 with respect to the virial temperature, despite the
cooling time is much shorter than the age of the cluster.  Again,
another evidence that some homogeneous process heats the gas.

\begin{figure}
\vskip 1.5cm
\begin{center}
{\bf see fig14.gif}
\end{center}
\vskip 2cm
\caption{From left to right: AGN activity creating cavities in the ICM
of the Perseus cluster (Fabian et al. 2002; 2003); cold fronts in
Abell 2142 (Markevitch et al. 2000); an ongoing massive merger in 1E
0657--56, the {\sl bullet cluster} (Markevitch et al. 2004).}
\label{bullet}
\end{figure}

Today, we see clearly that the Chandra and XMM satellites changed our
perspective of clusters of galaxies.  If in the ROSAT era the main
goal was to find as many clusters as possible with the aim of
constraining cosmology, in the Chandra/XMM era the goal is to observe
with much better spatial and spectral resolution the clusters
previously discovered.  The physics of the ICM is much more complex
than expected and this forces us to reconsider all the relations
between the X--ray observables and the dynamical mass.  This aspect
may cast some doubts on the use of X-ray clusters of galaxies as
cosmological tools.  One can also think to reverse the argument: the
physics of the ICM is much more interesting, so let's investigate the
evolutionary properties of clusters to understand the effects of
feedback processes onto the ICM, and don't worry about cosmology.

In my view, the investigation of cosmology and of the ICM physics must
proceed together.  Actually, this is what is happening: if you go
through the literature in the last six years, you discover indeed that
there is still a strong interest in cosmological tests with clusters,
which is supported by a growing amount of works on the ICM.  It must
be noticed in addition, that understanding the problem of the
non--gravitational heating of the ICM by energetic feedback from star
formation or nuclear activity, is a key issue in cosmic structure
formation.  Actually, feedback is the holy grail of structure
formation today!  If you go to a conference on galaxies, clusters, or
anything on cosmic structure formation, you will hear everywhere the
word ``feedback''.  So, rather than saying that clusters became less
interesting in a cosmological perspective, I prefer to say that
clusters became even more important to understand both structure
formation and cosmology.

\subsection{A simpler cosmological test}

There is not enough space here to describe the most recent progress in
the understanding of the ICM thermodynamics.  However, I want to
mention another cosmological test that appears to be simpler than that
discussed so far.  Instead of relying on the knowledge of the dynamics
of clusters, we can focus on a much simple quantity: the baryonic
fraction $f_B$.  We simply need to measure the total mass, and count
all the baryons in the form of stars and ICM.  From semianalytical
models and numerical simulations, we expect that the physics of the
ICM does not affect $f_B$ if measured at a radius where the gravity
dominates; therefore, it should be close to the cosmic value
$\Omega_B/\Omega_0$.  In other words, the baryons are allowed to
behave wildly and decouple from the dark matter distribution in high
density regions, but on large scales they are not displaced
differently from dark matter.  The virial radius is expected, then, to
include a closed region where the average composition does not change
during the evolution of the cluster.  It is straightforward to see
that the measure of $f_B$ and the knowledge of $\Omega_B$ from
nucleosinthesis or from the CMB (Spergel et al. 2003) gives a
straightforward measure of $\Omega_0$ (see White et al. 1993).

But this is not all: for the same reasons, the baryonic fraction
should not evolve with redshift. However, the actual measure of $f_B$
does depend on the angular distance.  The mass of baryons is recovered
by measuring the flux and by knowing the physical size of the cluster.
The relation between the messured flux $S_X$ and the mass of gas reads
as:

\begin{equation}
S_X = L_X(1+z)^{-4}/(4\pi d^2_{ang}) \propto M^2_{gas} \theta^{-3}_c
d^{-3}_{ang} / d^2_{ang}\, .
\end{equation}

\begin{figure}
\centering \includegraphics[height=5cm]{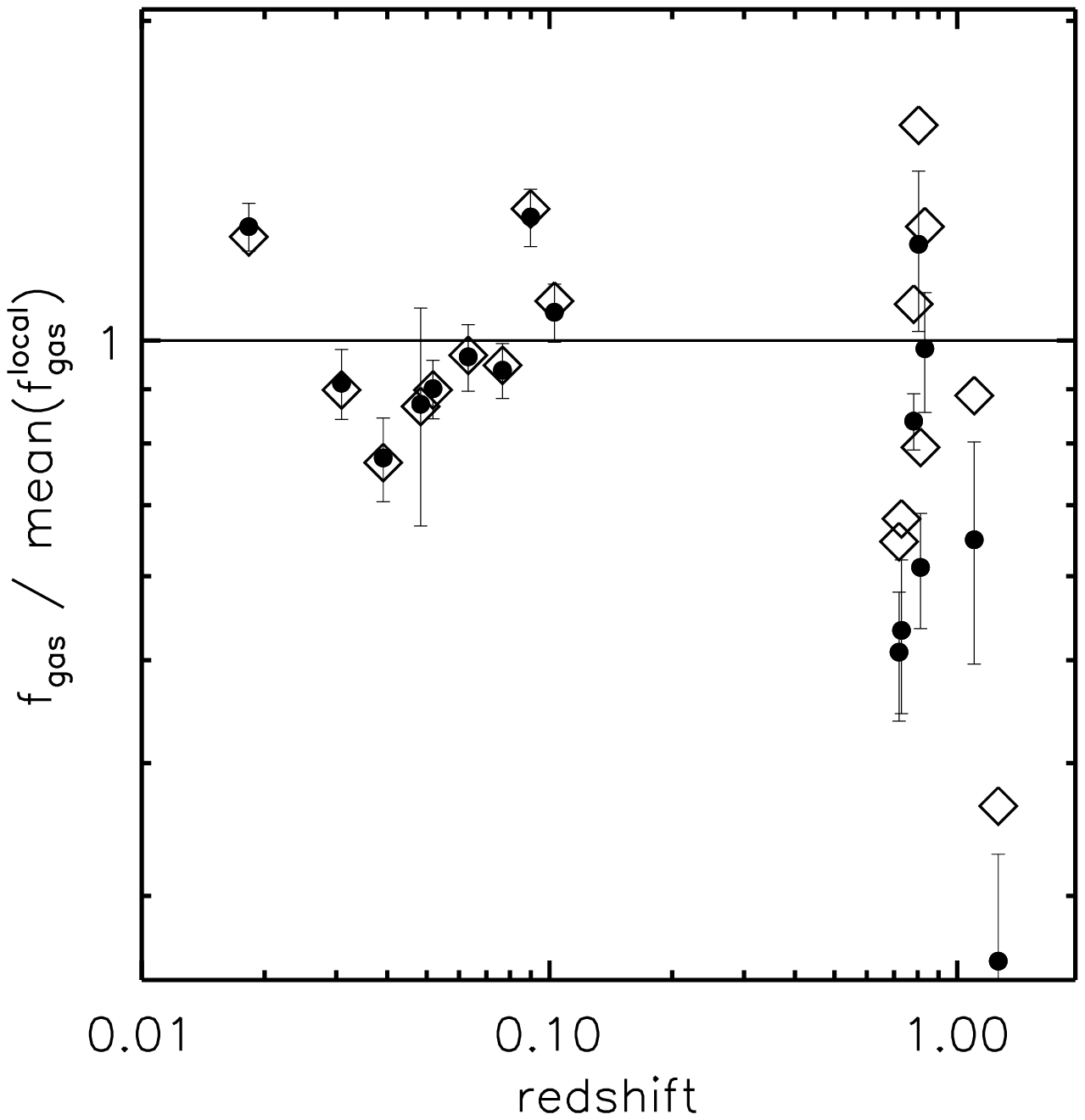}
\centering \includegraphics[height=5cm]{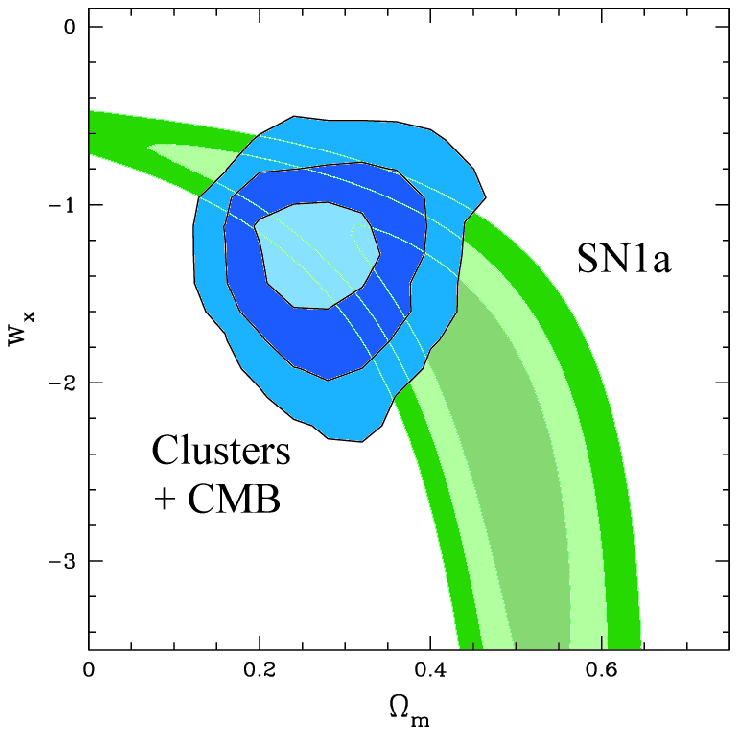}
%
\caption{Left: $f_B$ measured for a sample of high--z clusters in an
EdS cosmology (dots) and in a flat $\Lambda$ universe (empty diamonds,
from Ettori et al. 2003).  Right: constraints in the $w$--$\Omega_0$
plane obtained by combaining baryonic fraction in clusters and CMB
 (Allen et al. 2004).}
\label{ettori}
\end{figure}

\noindent
On the other hand, the total mass depends on the angular distance as
$M_{tot} \propto \theta_c d_{ang}$.  It follows that $f_B =
M_{gas}/M_{tot} \propto d_{ang}^{3/2}$.  Thus, we have two advantages
here: the value of the baryon density gives $\Omega_0$, while its
apparent evolution is depending on the cosmological parameters through
$d_{ang}$.  Therefore, the cosmological test consists in requiring no
evolution in the observed $f_B$.  Any apparent evolution in
the baryonic fraction is the smoking gun of wrong cosmological
parameters.  It is important to perform this test on a redshift range
as wide as possible (see Figure \ref{ettori}, left).

This is not a dynamical test, but rather a geometrical test, and it is
more sensitive to $\Omega_\Lambda$ (see Allen et al. 2004).  However,
we notice that the scatter in the baryonic fraction from cluster to
cluster is somewhat larger than we would like, given the starting
assumption of a universal value for $f_B$ for all clusters at all
epochs.  This is probably due to the fact that the dynamical masses
and the baryonic fraction measures are still affected by complexities
in the ICM physics (see Hallman et al. 2005).  However, this kind of
test is very promising, and it becomes very powerful when combined
with CMB or SNeIa test, as shown in Figure \ref{ettori} (right).

\subsection{Future prospects for precision cosmology with clusters}

\begin{figure}
\centering \includegraphics[height=7cm]{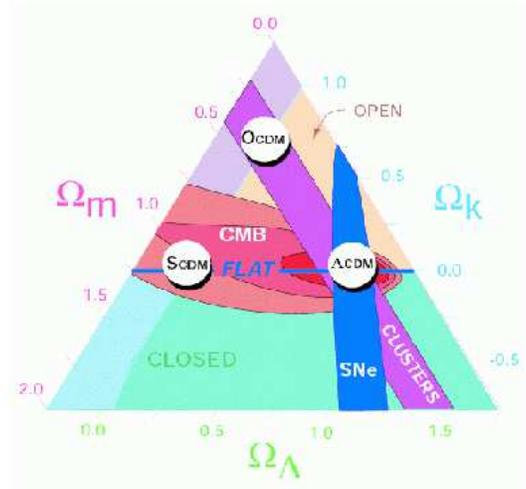}
%
\caption{The cosmic triangle (Steinhardt 2003, see
http://wwwphy.princeton.edu/~steinh/).  The complementarity of the
three classic cosmological tests is clearly shown. }
\label{triangle}
\end{figure}

We are approaching the end of our brief introduction to cosmological
tests with clusters of galaxies.  A clear way to summarize it, is the
cosmic triangle shown in Figure \ref{triangle}.  Each side represents
one of the three main parameters: the mass density, the cosmological
constant, and the curvature.  Contours levels perpendicular to one
side mean that a particular test is efficient in constraining that
parameter.  Cosmological tests based on clusters are mostly sensitive
to $\Omega_0$, while geometrical tests like CMB and SNeIa are more
sensitive to the curvature and $\Omega_\Lambda$.  Roughly speaking,
CMB can constrain $\Omega_0+\Omega_\Lambda$, while SNeIa
$\Omega_0-\Omega_\Lambda$, mainly beacuse of the different redshift
range, 0.5--2 for SNeIa and ~1000 for CMB.  Obviously, the combination
of the three tests is very powerful, but its application requires a
good understanding of all the different systematics.

This picture is still valid after recent observations by Chandra and
XMM showed that the physics of clusters is more complicated than
expected.  The key questions on the future of precision cosmology with
clusters of galaxies are: do we need a new, large, all--sky survey of
clusters?  Or should we first understand better the physics of the
ICM?  Therefore, which is the best instrument we should build next?  I
think that the best answer is that a new, medium--depth all--sky
survey of clusters is needed for both aspects.  First, a large survey
can help in obtaining strong constraints on the cosmological
parameters, providing at the same time large samples to investigate
the relationship between X--ray observables and the dynamical masses.
A second crucial aspect, is that a large survey would discover new
clusters, especially at high redshift.  This is mandatory to provide
targets for the future X--ray missions, which will provide sensitive,
narrow--field instruments to investigate the physics of the ICM.
Without a new wide survey, we will run out of clusters to observe!
Several proposals of medium--size mission have been circulated so far,
but at present there are no planned large--area surveys of the X--ray
sky.  The future of X--ray cluster astrophysics largely depends on
this.

\section{What to bring home}

At the end of this introduction, we should be aware that clusters of
galaxies constitute a cosmological tools to significantly constrain
$\Omega_0$ and the spectrum of primordial fluctuations, through tests
based on dynamics or on geometry.  X--ray observations offer the best
tool to measure mass and collect complete sample of clusters.  Main
results points towards a flat $\Lambda$--dominated Universe ($\Omega
\sim 0.3$ and $\Omega_\Lambda \sim 0.7$, or $w=-1$) and a
normalization of the fluctuations power spectrum consistent with that
measured from CMB for a CDM Universe ($\sigma_8 \simeq 0.8$).

If someone wants to start the business of cosmological tests with
clusters, she/he just needs basic programming skills to put in a
simple code all the formulae we discussed, and a good X--ray observer
among the collaborators, in order to have access to a well
defined, complete survey of clusters.  However, one must know that
this game was played a lot starting from the 90's, when it was
realized that clusters constitute one of the most powerful
cosmological tools.  At present, in 2006, most of the best X--ray
clusters surveys have been exploited in this sense.  Therefore, if you
want to start the business, you better have something smart in mind,
mainly a way to deal with any possible systematics or with a better
treatment of the effects of the poorly known thermodynamics of the
ICM.

However, a noticeable contribution would be given by supporting the
scientific case of future X--ray missions to obtain new data, in the
form of a wide and complete sample of clusters.  Larger samples
indeed, will allow to study at the same time the thermodynamics of the
ICM and the evolution of clusters as a population.  Finally, one
should not have the feeling that the physics of the ICM is now the hot
topic at the expenses of precision cosmology, which should rely only
on tests based on SNeIa and CMB.  As a general comment, I would like
to stress that clusters are probing a different cosmic epoch with
respect to CMB, and a different physics with respect to SNeIa,
therefore they will always be a complementary and useful test for
cosmology.  The complex ICM physics, insted of being an obstacle, must
be seen as a further oppurtunity to learn about structure formation in
the Universe.

%
%
%

%
%



\printindex
\end{document}